\documentclass[prl,aps,twocolumn,superscriptaddress,nofootinbib]{revtex4}

\usepackage{epsfig}
\usepackage{amsfonts}
\usepackage{latexsym}
\usepackage{amsmath}
\usepackage{amssymb}
\usepackage{slashed}

\usepackage{graphicx}
\usepackage{hyperref}

\def\half{{1\over 2}}
\numberwithin{equation}{section}

\newcommand{\bea}{\begin{eqnarray}}
\newcommand{\eea}{\end{eqnarray}}
\newcommand{\be}{\begin{equation}}
\newcommand{\ee}{\end{equation}}
\newcommand{\ba}{\begin{align}}
\newcommand{\ea}{\end{align}}

\newcommand{\DD}{\slashed{D}}
\newcommand{\hDD}{\hat{D}}

\def\half{{1 \over 2}}

\def\Or[#1]{{\text{O}}\left({#1}\right)}
\def\dotl[#1,#2]{\left\langle #1, #2 \right\rangle}
\def\dotlb[#1,#2]{[ #1, #2 ]}
\def\dotp[#1,#2]{(#1) \cdot (#2)}
\def\aff[#1,#2]{\hat{#1}(#2)}
\def\n4sym{{\cal N}=4 SYM}
\def\>{\rangle}
\def\<{\langle}
\def\weight[#1,#2,#3]{\{(#1),#2,#3\}}
\def\ads[#1]{$\text{AdS}_{#1}$}

  \makeatletter
  \let\over=\@@over \let\overwithdelims=\@@overwithdelims
  \let\atop=\@@atop \let\atopwithdelims=\@@atopwithdelims
  \let\above=\@@above \let\abovewithdelims=\@@abovewithdelims

\newcommand{\vn}{{\vec{n}}}

\usepackage{xcolor}

\begin{document}

\preprint{}

\title{One-loop effective actions and 2D hydrodynamics with anomalies}

\author{Gim Seng Ng} \affiliation{Departments of Physics, McGill University, Montreal, H3A 2T8, Canada }
\author{Piotr Sur\'owka} \affiliation{Center for the Fundamental Laws of Nature, Harvard University, Cambridge, MA 02138, USA}

\begin{abstract}
We revisit the study of a 2D quantum field theory in the hydrodynamic regime and develop a formalism based on Euclidean one-loop partition functions that is suitable to analyze transport properties due to gauge and gravitational anomalies. To do so, we generalize the method of a modified Dirac operator developed for zero-temperature anomalies to finite temperature, chemical potentials and rotations.
\end{abstract}

\maketitle
\emph{Introduction.}---Quantum anomalies play an important role in the hydrodynamic regime, leading to new transport properties \cite{Son:2009tf,Landsteiner:2011cp}. This may have consequences in various physical systems such as heavy-ion collisions, cosmology, and condensed matter systems \cite{kharzeev2013strongly}. Therefore a detailed theoretical understanding of anomalies in relativistic fluids is necessary and has led to numerous recent developments in the field. Much of the work has been based on the consistency of hydrodynamic expansions, including second law of thermodynamics, or semiclassical kinetic theory analysis \cite{Loganayagam:2012pz,Loganayagam:2012zg,Jensen:2012kj,Jensen:2013kka,Jensen:2013rga,Son:2012wh,Stephanov:2012ki}. In addition, the necessity to include anomalies in fluid dynamics is supported by AdS/CFT arguments \cite{Erdmenger:2008rm,Banerjee:2008th,Amado:2011zx,Azeyanagi:2013xea}.

A natural framework to study anomalies at zero temperature utilizes heat kernels or zeta-function regularizations \cite{Bertlmann,Fursaev:2011zz}. The advantage of these techniques comes from the fact that they can be used to describe different types of anomalies, including continuous and discrete anomalies. However, previous attempts \cite{BoschiFilho:1991ah,BoschiFilho:1991xz} to capture anomalous contribution at finite temperature (but without rotations) did not give any non-zero result. Using hydrodynamic reasoning and state/operator counting in free massless theories, however, we now know that in order to get a non-zero contribution we need to consider rotating fluids \cite{Loganayagam:2012pz,Loganayagam:2012zg}.

The aim of this paper is to show that in the Euclidean path-integral method the spectrum of an appropriate ``chiral'' operator \cite{AlvarezGaume:1983cs} at finite temperature and density for rotating fluids indeed gives the required chiral-half of the results. This paper concentrates on a (1+1)-dimensional rotating fluid (by which we mean a Lorentz-boosted fluid), though we expect and hope to apply the method to higher-dimensional fluids in the presence of magnetic fields.

Let us consider a Dirac-type operator $D$ on a Euclidean torus $T^2$, upon which the Euclidean path-integral is used to compute the canonical partition function at finite temperature $\beta$, chemical potential $\mu$ and rotation $\Omega$. The one-loop approximation to the partition function is:
\be
Z[\beta,\mu,\Omega]\approx e^{-S_{cl}}\det [ D] \,,
\ee where the classical contribution is denoted as $e^{-S_{cl}}$.
Our aim is to calculate the one-loop effective action:
\be
W[\beta,\mu,\Omega]\equiv\ln Z[\beta,\mu,\Omega]+S_{cl}\approx \ln \det[D].
\ee
It describes quantum effects in the presence of background fields
in the one-loop approximation of quantum field theories.

In this Letter, we show how to isolate the chiral-contribution to the thermodynamics of a (1+1)-dimensional field theory with global anomalies living on the manifold $T^2_{(\beta,\Omega)}$ with coordinates $(t,x)$. The thermal circle's periodicity is $(t,x)\sim(t+\beta,
 x + \beta \Omega)$ while the compact spatial circle's periodicity is given by $(t,x)\sim(t,x+1)$.  The complex parameter of the torus is then $\tau\equiv \tau_1+i \tau_2\equiv \Omega\beta+ i {\beta }/{(2\pi R)}$, where $R$ is the radius of spatial circle. We are interested in studying the one-loop effective action $W[\beta,\mu,\Omega]$ in the high temperature limit where $\Omega \beta\ll 1$ and $\beta/R \ll 1$ or, equivalently, $\tau\rightarrow 0$. A method developed to account for the contribution due to chiral and Lorenz anomalies at finite temperature and the interpretation in terms of anomalous hydrodynamics is the central result of this Letter.

\emph{2D Free Dirac Fermion}---To warm up and provide a starting point for the subsequent inclusion of anomalies, let us first review the computation of the partition function of a non-anomalous, massless, free Dirac operator
 $D={(\beta/(2\pi))} (i\DD)$.
It can be obtained through
\be \label{eq:diracdet0}
\det[D]=\sqrt{\det [D^2]}\,.
\ee One trick to evaluate $\det[D^2]$ is to find Weyl spinors $\psi_{\vn}^\pm$ (i.e. $\Gamma_5 \psi^\pm_{\vn} = \pm \psi^\pm_{\vn}$) which furthermore satisfy
\be\label{eq:spinoreigencond}
D\psi^\pm_{\vn}= \lambda^{\pm}_\vn \psi^\mp_{\vn}\, ,
\ee such that $\psi^\pm_{\vn}$ diagonalizes $D^2$:
\be
D^2 \psi^\pm_{\vn} = \lambda_\vn^2 \psi^\pm_{\vn},\quad
\lambda^2_\vn\equiv \lambda^+_\vn \lambda^-_\vn\,.
\ee
As a result we obtain
\be
\det [D^2]
= \prod_{\vn} \lambda_\vn^+ \lambda_\vn^-\,,
\ee and thus
\be \label{eq:diracdet}
\det[D]=\sqrt{\prod_{\vn}( \lambda^+_{\vn} \lambda^-_{\vn})}\,.
\ee

In this work, we will always impose the $(A,A)$-boundary condition (where $A$ means anti-periodic while the first (second) slot indicates the thermal (spatial) circle):
\be\label{eq:bcAA}
\psi_{\vn}^\pm(t,x)= -\psi_{\vn}^\pm(t,x+1)=-\psi_{\vn}^\pm(t+\beta,x+\beta  \Omega) \, .
\ee
A basis for such Weyl spinors satisfying Eq.~(\ref{eq:spinoreigencond}) and Eq.~(\ref{eq:bcAA}) is provided by
\bea
&&\psi^\pm_\vn(t,x) \nonumber\\
&\propto& \exp{[
+2 \pi i (n_1 +1/2) {t}/{\beta}
+2\pi i (n_2+1/2)( x-t \Omega)]}
 u_\pm\,,\nonumber\\
\eea where
$
 \Gamma_5 u_\pm = \pm u_\pm$ and
$\vn\in {\mathbb Z}^2.$
In our conventions,
\be
i\DD=
 \left( \begin{array}{cc}
0 & -\partial_{t}+ i(2\pi R)^{-1}\partial_x\  \\
\partial_{t}+ i(2\pi R)^{-1}\partial_x & 0
\end{array} \right)
\,.
\ee
Applying $(\beta/(2\pi))i\DD$ to $\psi_\vn^\pm$, we obtain
\be
\label{eq:lambdapORlambdam}
 \lambda^+_{\vn} =i\left[
\left(n_1+\half\right) -\left(n_2+\half\right) \bar{\tau}
 \right]
 ,\quad
 \lambda^-_{\vn} = (\lambda^+_{\vn})^*\,.
 \ee
\be
\label{eq:lambdaplambdam}
 \lambda^+_{\vn} \lambda^-_{\vn}
=
\left|\left(n_1+\half\right)-\left(n_2+\half\right) \tau  \right|^2\,.
\ee
In evaluating the infinite product of the eigenvalues, we use the identity
\be
\prod_{n_1=-\infty}^\infty \left(n_1+\half+a\right)
=-2i\cos{(\pi a)}
\ee and the Hurwitz zeta function regularization of
\be\label{eq:prescriptionII}
\left[
\prod_{n=0}^{\infty} q^{-\half\left(n+\half\pm a\right)}
\right]
=q^{-\half \zeta(-1,1/2\pm a)}
=q^{-1/48+a^2/4}.
\ee
The result for the one-loop effective action is
\begin{eqnarray}
W[\tau,\bar{\tau}]
&=&\half \log\left|  q^{-1/24}  \prod^\infty_{n = 1}\left(1+q^{n-\half }
\right)^2\right|^2\nonumber\\
&=&\half\log\left| \frac{\theta(\tau)}{\eta(\tau)} \right|^2\,,\quad
q\equiv e^{2\pi i\tau}\,,
\end{eqnarray} where the $\eta$ and $\theta$ functions are defined as
\bea
\eta(\tau)&\equiv& q^{1/24} \prod_{n=1}^\infty \left(1-q^n\right)\, , \nonumber\\
\theta(\tau)&\equiv &q^{-1/24} \eta(\tau) \prod_{n=1}^\infty \left(1+ q^{n-1/2}\right)^2\,.
\eea Using the modular properties of these functions
\be
\frac{\theta(\tau)}{\eta(\tau)}
=\frac{\theta(-1/\tau)}{\eta(-1/\tau)}\,,
\ee  and their asymptotic expansions, we obtain as $\tau \rightarrow 0$
\be
\lim_{\tau\rightarrow 0}
\frac{\theta(\tau)}{\eta(\tau)}
= \exp\left[{+2\pi i \frac{1}{24\tau}}\right]
\ee implying the well-known Cardy's formula for a 2D massless Dirac fermion
\be
\lim_{\tau\rightarrow 0}W[\tau,\bar{\tau}]
=+2\pi i \frac{(1/2)}{24\tau}
-2\pi i \frac{(1/2)}{24\bar{\tau}}
\ee consistent with the fact that a 2D massless Dirac fermion is equivalent to one left moving Weyl fermion and one right moving Weyl fermion with central charges $c_L=c_R=1/2$.
We note that although in some sense a Dirac fermion is equivalent to two Weyl fermions, the chiral contribution (i.e. contributions proportional to $c_L-c_R$) cannot be captured by taking an appropriate half of the Dirac's answer \cite{Loganayagam:2012zg}.
This is related to the fact that the anomalous contribution to the effective action is purely imaginary \cite{AlvarezGaume:1983cs}, while the above answer is purely {\it real}.
A way to overcome this issue and to isolate the Weyl fermion's contribution in the Euclidean path integral has been proposed by \cite{AlvarezGaume:1983cs} and will be the focus of the next sections.

\emph{Chiral chemical potential}--- In this section, we will show how to appropriately calculate a determinant which captures the contribution of a single Weyl fermion. Let us first introduce a non-trivial chemical potential. It can be done either by modifying the Dirac operator in a chiral way or equivalently by imposing
boundary conditions for the basis spinors differently depending on their chirality:
\be\label{eq:bcchiralcoupling}
\psi_{\vn}^\pm(t,x)= -\psi_{\vn}^\pm(t,x+1)=-e^{i\beta  \mu_{\mp}}\psi_{\vn}^\pm(t+\beta,x+\beta  \Omega) .
\ee Then, it is easy to see that Eq.~(\ref{eq:lambdaplambdam}) will be modified in the following way:
\bea
\label{eq:lambdaplambdam2}
 \lambda^+_{\vn} \lambda^-_{\vn}
&=&
\left[\left(n_1+\half+\nu_-\right)-\left(n_2+\half\right) \bar{\tau} \right]\nonumber\\
&&\times
\left[\left(n_1+\half+\nu_+\right)-\left(n_2+\half\right) {\tau }\right]\,,
\eea where $i\beta  \mu_\pm\equiv {2\pi  i\nu_{\pm}}$ and $\nu_\pm$'s are real. Similar computations as before now yield
\be
W[\tau,\nu_+;\bar{\tau},\nu_-]
=\half \log \left[
\frac{{\theta(\nu_+,\tau)}}{{\eta(\tau)} }
\frac{\overline{\theta(\nu_-,\tau)}}{\overline{\eta(\tau)} }
\right]
\,,
\ee where $z_\pm=e^{2\pi i \nu_\pm}$ and
\be
\theta(\nu_\pm,\tau)\equiv \prod_{n=1}^\infty (1-q^n)(1+z_\pm q^{n-1/2})(1+z_\pm^{-1} q^{n-1/2})\,.
\ee Note that $\theta(\tau)=\theta(0,\tau)$.
The $\theta(\nu,\tau)$ function  has the following modular transformation:
\be
\frac{\theta(\nu,\tau)}{\eta(\tau)}
=e^{-\pi i \nu^2 /\tau}~\frac{\theta(\nu/\tau,-1/\tau)}{\eta(-1/\tau)} \, ,
\ee which has the small $\tau$ behavior of the form
\be
\lim_{\tau\rightarrow0}
\frac{\theta(\nu,\tau)}{\eta(\tau)}=\exp{\left[
2\pi i \left(\frac{1}{24 \tau}- \frac{\nu^2 }{2\tau}\right)
\right]} \, .
\ee This gives
\bea
&&\lim_{\tau\rightarrow 0}W[\tau,\nu_+;\bar{\tau},\nu_-]\nonumber\\
&=&
2\pi i \left[\frac{(1/2)}{24\tau}
- \frac{(1/4)\nu_+^2 }{\tau}
\right]
-2\pi i \left[\frac{(1/2)}{24\bar{\tau}}
- \frac{(1/4)\nu_-^2 }{\bar{\tau}}
\right]\,,
\nonumber\\
\eea which reproduces the Cardy's formula (with chemical potentials) with central charges $c_L=c_R=1/2$ and Kac-Moody levels $k_L=k_R=1/4$.

Tuning $\nu_-=0$ and $\nu_+ \ne 0$ (or vice versa), we could isolate the chiral-half of the answer, which probes the $k_L$ or $k_R$ part independently.
The above result is not accidental
but follows from the identification of of chiral determinants with a determinant of a modified Dirac-type operator which acts on Dirac spinors \cite{AlvarezGaume:1983cs}. The equivalence between the ``chiral boundary conditions'' introduced above and the modified Dirac-type operator method of \cite{AlvarezGaume:1983cs} will be explained below.

In the construction of \cite{AlvarezGaume:1983cs}, one naively starts by trying to study the eigenvalue problem of the operator
\be
i{\DD}_+=i\Gamma^\mu\left(\partial_\mu + A_\mu\right) P_+,\quad P_\pm\equiv \half\left(1\pm\Gamma_5\right),
\ee and identifies the path-integral over a single Weyl mode as the determinant of such an operator.
However, since this operator maps the $P_+=1$ subspace to the $P_+=0$ subspace, the eigenvalue problem for $i{\DD}_+$  is not well-defined.
Instead, one studies the modified operator
\be
\hat{D}\equiv  i\Gamma^\mu\left(
\partial_\mu
+A_\mu P_+
\right)= i\begin{pmatrix}
 0 & \slashed{\partial} _-\\
 \slashed{D}_+ & 0\\
 \end{pmatrix}\, ,
\ee where the background field $A_\mu$ is coupled {\it only} to the chiral half of the operator. Since $\hat{D}$ acts on Dirac fermions, it has twice the number of degrees of freedom as the operator of Weyl fermions. But this doubling would affect only the overall normalization and not the effective action (as a functional of the background field).

Furthermore, following \cite{AlvarezGaume:1983cs}, the $\det[\hat{D}]$ can be computed by realizing that
\be
\det[\hat{D}] = \det [(i\slashed{\partial} _{-})(i  \slashed{D}_+)]\,.
\ee If we found Dirac spinors $\psi_\vn^+$ such that
\be
(i \slashed{\partial} _{-} )(i\slashed{D}_+) \psi _\vn^+= \lambda_ \vn ^2 \psi _\vn^+ \, ,
\ee then
\be
\det(\hat{D})=\prod_{\vn} \lambda_ \vn ^2\,\, .
\ee A particular convenient choice will be to find Weyl spinors $\psi^\pm_{\vn}$ with the properties that
\be\label{eq:lambdaplusminusv2}
i\slashed{D}_+ \psi_{\vn}^+ = \lambda_\vn^+ \psi_{\vn}^-,\quad
i\slashed{\partial}_- \psi_{\vn}^- = \lambda_\vn^- \psi_{\vn}^+,
\ee such that $\lambda^2_\vn=\lambda^+_\vn \lambda^-_\vn$. The determinant now reads\footnote{We note that a different choice of basis may change the the $\lambda _{n}^-$ and $ \lambda _{n}^+$ (as defined in Eq.~(\ref{eq:lambdaplusminusv2})) but the products $\lambda _{n}^- \lambda _{n}^+$ remain unchanged \cite{Sachs:1995dm}, i.e.
\be
\lambda^2_\vn=\lambda_\vn^+ \lambda_\vn^-
\ee is independent of the choice of basis for the $\psi^\pm_{\vn}$.
}
\be
\det(\hat{D})=\prod_{\vn} (\lambda_\vn^+ \lambda_\vn^-)\, .
\ee

Let us now compute the determinant of $\hat{D}$. Restricting to $A_a dx^a=A_0 dt$, we have
\be
\hat{D}= \begin{pmatrix}
 0 & -(\partial _t +A_0)+i(2\pi R)^{-1}\partial _x\\
 \partial _t + i(2\pi R)^{-1}\partial _x & 0\\
 \end{pmatrix} \,.
\ee
Thus, the effect of introducing the chiral coupling to the gauge field is equivalent to shifting the ``upper-right" part of the operator by $\partial_t\rightarrow \partial_t + A_0$.
This implies that the $\lambda_\vn^-$ will get shifted appropriately while $\lambda_\vn^+$ remains unchanged:
\bea
 \lambda^+_\vn&=&i\left[
\left(
n_1+\half\right)
-\left(n_2+\half \right)\bar{\tau}
\right],\quad \\
\lambda^-_\vn&=&-i\left[
\left(
n_1+\half+\nu_+\right)
-\left(n_2+\half\right)\tau
\right]\,,
\eea
 once we have identified
\be\label{eq:A0munu}
A_0
 =i \mu_+
=i\frac{\nu_+}{R \tau_2 }\, .
\ee Note that since $A_a dx^a=A_0 dt$ is defined such that $t$ is the Euclidean time, the $A_0$ is purely imaginary for real $\mu_+$.

The above computation shows that the chiral coupling is implemented by sending
$n_1\rightarrow
n_1+\nu_+$ in $\lambda_\vn^-$. This is equivalent to not modifying the operator $D$ but instead changing the boundary conditions for the $\psi^+_\vn$ modes according to Eq.~(\ref{eq:bcchiralcoupling}) with $\mu_-=0$ but $\mu_+=\nu_+/(R\tau_2)$. This argument demonstrates that one can interchangeably implement chiral couplings either by: i) imposing chiral boundary conditions while using $D$ or ii) using the chiral operator $\hat{D}$ instead of $D$ together with non-chiral boundary conditions. We will use this property to calculate the chiral contribution of rotation to our system.

\emph{Chiral coupling to rotation}---Motivated by the equivalence presented in the previous section, we shall now investigate what happens in the case of zero chemical potentials, but instead we impose the ``chiral boundary condition" for the rotation.
More precisely, we require
\bea
\psi_{\vn}^\pm(t,x)&=& -\psi_{\vn}^\pm(t,x+1)\nonumber\\
\psi_{\vn}^-(t,x)&=&-\psi_{\vn}^-(t+\beta,x+\beta  \Omega) \nonumber\\
\psi_{\vn}^+(t,x)&=&-\psi_{\vn}^+(t+\beta,x) \, ,
\eea  i.e. the ``twist" induced by the rotation is imposed only on one of the Weyl spinors.
Similar to the computations in the chemical potential case in the previous section, it is straightforward to show that the above ``chiral boundary condition'' for the rotation is equivalent to the modification of the operator to ``chiral-couple'' rotation to the vielbeins. At zero temperature, such modifications of the operator or boundary conditions exactly reproduce what one expects from Lorentz anomalies. This was proven in \cite{Fursaev:2011zz}.
At finite temperature and rotations, we will now show that rotations can be incorporated in a similar manner and the outcome agrees with the standard Cardy's formula.

In analogy with the previous sections we calculate appropriate eigenvalues
\bea\label{eq:chiralrotationeigenvalues}
\lambda^+_\vn&=&i\left[
\left(
n_1+\half\right)
-\left(n_2+\half \right)\bar{\kappa}
\right],\quad\\
\lambda^-_\vn&=&-i\left[
\left(
n_1+\half\right)
-\left(n_2+\half\right)\tau
\right]\,,
\eea
where $\kappa \equiv  i \beta/(2\pi R)$, $\kappa_2= \beta/(2\pi R)$ and $\tau\equiv \Omega\beta+ i \beta/(2\pi R)$.
The rest of the computation proceeds as before:
\be
\det[ \hDD ]
=
\frac{
\theta(\tau)
}{\eta(\tau)}
\frac{
\overline{\theta(\kappa)}
}{\overline{\eta(\kappa)}}\,\, \Rightarrow\,\,
W[\tau,\bar{\tau}]
=\half\log\left[
\frac{
\theta(\tau)
}{\eta(\tau)}
\frac{
\overline{\theta(\kappa)}
}{\overline{\eta(\kappa)}}
\right]\, ,
\ee
where
$q\equiv e^{2\pi i \tau}$ and
$p\equiv e^{2\pi i \kappa}$.
Using the standard modular transformations and the asymptotic expansions of the theta and eta functions, we obtain
\be
\lim_{\beta\rightarrow 0}W[\tau,\kappa]
=\half
\left[
2\pi i \left(\frac{1}{24 \tau}\right)
-2\pi i \left(\frac{1}{24 \kappa}\right)
\right]\,.
\ee

To isolate the chiral or anomalous contributions, we should now take the {\it imaginary} part of the $W$ \cite{AlvarezGaume:1983cs}. Thus, we define
\be
W_{\text{anom}}[\beta,\Omega]\equiv  \text{Im}\left[ W[\beta,\Omega] \right]
\ee leading to
\bea
\lim_{\beta\rightarrow 0}W_{\text{anom}}[T,\Omega]
&=&\frac{1}{T}  \left[\frac{(2\pi R)(2\pi R \Omega)}{ (1-(2\pi R\Omega )^2)} \right] 2\pi \left[\frac{(1/2)}{24}T^2\right]\, ,\nonumber\\
\eea which is just the imaginary part of the Cardy's formula with $c_R=1/2$ and $c_L=0$. We see that the {\it imaginary} part of $\det[\hat{D}]$ automatically computes the anomalous contribution to the effective action (and hence all the thermodynamical quantities) in the high-temperature limit.

\emph{Conclusions}---In this Letter, we developed a method to deal with finite-temperature anomalies using one-loop determinants of an appropriate operator in two dimensions. This method can be generalized to higher dimensions and has many potential applications. It would be interesting to investigate its analogue in odd dimensional field theories in the study of discrete anomalies at finite temperature. Another interesting direction would be to isolate the contributions of anomalies to entanglement entropy with this technique. Finally we believe that the generalization to all even dimensions can provide yet another perspective on the anomaly polynomial "replacement rule", which captures anomalous contributions to the partition function at finite $T$ and $\mu$ \cite{Loganayagam:2012pz,Loganayagam:2012zg,Jensen:2012kj,Jensen:2013rga,Jensen:2013kka}.

We thank T.~Dumitrescu, G.~Dunne, K. Jensen and R.~Loganayagam for useful discussions and correspondence.  PS is grateful to the Mainz Institute for Theoretical Physics (MITP) and Caltech for their hospitality and partial support during the completion of this work. GSN was supported by an NSERC Discovery Grant. PS was supported by a Marie Curie International Outgoing Fellowship, grant number PIOF-GA-2011-300528.
\bibliographystyle{apsrev4-1}

\bibliography{anomHK-bib}

\end{document}